\documentclass[global,twocolumn]{svjour}

\usepackage{graphics}
\usepackage{float}

\journalname{Meteorology and Atmospheric Physics}

\begin{document}

\title{Detrended fluctuation  analysis of daily temperature records:
Geographic dependence over Australia}

\author{Andrea Kir\'aly \and Imre M. J\'anosi}
\institute{Department of Physics of Complex Systems,
E\"otv\"os University, P\'azm\'any P\'eter  s\'et\'any 1, H-1117 Budapest, Hungary. \\
Tel:+(36)-(1)-372-2878, fax:+(36)-(1)-372-2866, E-mail: janosi@lecso.elte.hu}

\date{Received: date / Revised version: date}
\maketitle

\begin{abstract}
Daily temperature anomaly records are analyzed (61 for Australia, 18 for Hungary) by
means of detrended fluctuation analysis. Positive long range asymptotic correlations
extending up to 5-10 years are detected for each case. Contrary to earlier claims, the
correlation exponent is not universal for continental stations. Interestingly, the dominant
factor is geographic latitude over Australia: the general tendency
is a decrease of correlation exponent with increasing distance from the equator.
This tendency is in a complete agreement with the results found by Tsonis et al.~(1999) for
500-hPa height anomalies in the northern hemisphere. The variance of fluctuations
exhibits an opposite trend, the larger is the distance from the equator, the larger
the amplitude of intrinsic fluctuations. The presence of Tropospheric Biennial Oscillation
is clearly identified for three stations at the north-eastern edge of the Australian continent.
\end{abstract}


\section{Introduction}
Sophisticated detection studies demonstrate that the mean global surface temperature
has increased by $\sim 0.06^{\circ}$C per decade in the 20th Century and by  $0.19^{\circ}$C
per decade between 1978 and 1998 (Houghton et al., 2001).
All projections of future change indicate that the warming is likely to continue, and
this conclusion holds regardless of the computer model used or the emission scenario applied
in the model (Zwiers, 2002). Besides an improving reproduction of current mean atmospheric state,
confidence in the simulation and prediction skills of global climate models requires
the reproduction of long-time correlation properties revealed by measurements.
A recent test by Govindan et al.~(2002) of seven state-of-the-art global  models
failed to reproduce the scaling behavior of  six long temperature records
by underestimating  the long range persistence of the atmosphere. Similar
discrepancies were detected already by Syroka and Toumi (2001a, 2001b).
Quite the contrary, Fraedrich and Blender (2003) demonstrated that long range
correlations can be reproduced by a numerical model if the coupling with
ocean is represented properly (although with fixed composition of greenhouse
gases).
In any case, direct comparison of local observations with the rather low resolution
general circulation models might have many pitfalls (Pielke et al., 2002),
therefore an extended analysis of measured asymptotic correlations is a prerequisite.

Various methods are used to characterize quantitatively the fluctuations and correlations
of high frequency meteorological data. Power density spectra
(Marple 1987; Percival and Walden 1993)
have been routinely computed for decades. Pelletier (1997) determined
the temperature spectra for hundreds of stations and ice core records and identified
power-law behavior for continental and maritime locations.
Koscielny-Bunde et al.~(1998) reported to observe a near universal exponent value
in the fluctuations of daily temperatures, however Weber and Talkner (2001)
found differences depending on the altitude of the meteorological station.
 Probably the largest related study up to now is presented by
 Eichner {\it et al.}~(2003)
 on temperature records of 95 stations all over the globe. They obtained
correlation exponent values in the range 0.55-0.9 centered strongly at around  0.65 over
continents, but systematically higher values for islands. The last observation is supported
by Monetti {\it et al.} (2003), where they detected much stronger persistence
for sea surface temperature than values for air over ground.
Tsonis et al.~(1999) demonstrated scale invariance in the variability
of 500-hPa height ano\-malies and found a general tendency for the correlation
exponent to decrease with increasing latitude.
In general, scaling can be identified asymptotically from the longest
available records only, shorter-time correlations are almost fully explained by using
first or second order linear autoregressive models (von Storch and Zwiers 1999; Kir\'aly
and J\'anosi 2002).

In this work we exploit the method of detrended fluctuation analysis (DFA) which has
proven useful in revealing the extent of long range correlations in diverse time series
(Peng et al.~1994, 1995). Talkner and Weber (2000), and
Heneghan and McDarby (2000) pointed out that DFA and traditional power spectra
provide equivalent characterizations of correlated stochastic signals,
with the essential difference that DFA can effectively filter out slow trends.
The situation is similar to the connection of power spectra
and autocorrelation functions manifested by the Wiener-Khintchin theorem:
the information content is mathematically  the same, but in many cases spectral density functions
are more sensitive and better exploratory tools for real data (von Storch and Zwiers 1999).

Our analysis  is based on a high-quality daily temperature data set for Australia
(Torok and Nicholls 1996; Trewin and Trevitt  1996).
The same collection was utilized in many
 studies to reveal an increased wheat yield due to recent climate trends
(Nicholls 1997),  changing
maximum and minimum temperature trends for the globe (Easterling et al. 1997),
Australian land surface temperature variability (Jones 1999),
trends in extreme daily rainfall and temperature in Southeast Asia (Manton et al. 2001),
the impact of  land cover change on the Australian near-surface climate
(Narisma and Pitman 2003), etc.
Data for 61 (out from 107) stations were selected according to
the quality of their climate record, in terms of site standards, homogeneity and
completeness of the series, and to provide the best possible spatial coverage
 (48 for the continent, 13 for islands). The average time span of the records
 was 45 years with a minimum of 22  and a maximum of 120 years.
We evaluated also historical records of daily mean
temperatures measured at 18 weather-stations in Hungary
from 1951 until  1989. Amplitude distributions, power spectra
and autocorrelation functions were analyzed earlier essentially for the same
data set by J\'anosi and Vattay (1992), asymptotic scaling
and stochastic modeling have  been described
by Kir\'aly and J\'anosi (2002).

Since the DFA method is relatively new, we provide its concise exposition in
Section 2. Section 3 gives the details of our results, Section 4 is devoted to
a short discussion of the relevance and consequences of the findings.

\section{Detrended fluctuation analysis}

The most important advantage of DFA over conventional methods
(e.g., autocorrelation-, spectral- and Hurst-ana\-lysis)
 is that it permits the detection of intrinsic self-similarity
embedded in a seemingly nonstationary time series. Following the
work of Peng et al.~ (1994, 1995), several theoretical studies
elucidated the  power and limitations of filtering out various trends
from synthetic data series (Heneghan and McDarby 2000; Talkner and Weber 2000;
Hu et al.~2001; Kantelhardt et al.~2001; Chen et al.~2002).

\begin{figure*}
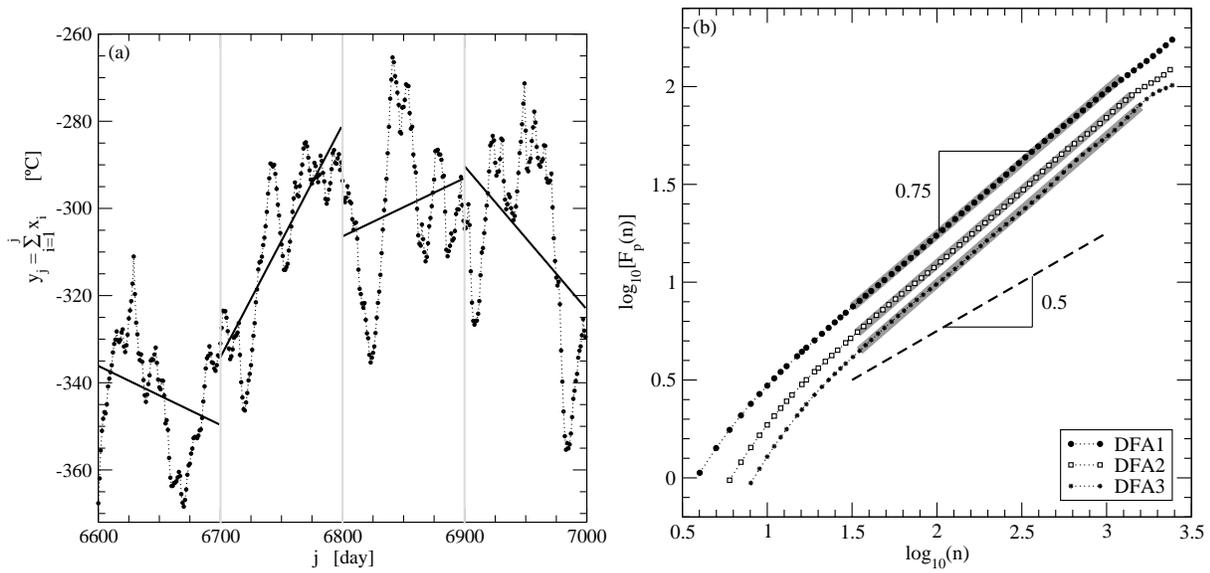

\begin{center}
\resizebox{0.45\textwidth}{!}{\includegraphics{KJ2fig1a.eps}}
\resizebox{0.44\textwidth}{!}{\includegraphics{KJ2fig1b.eps}}
\end{center}
\caption{(a) Local linear fits (solid line) for an integrated temperature anomaly
series $y_j$ [see (\ref{Eq:def})] divided into segments of length $n=100$.
(b) DFA results for a daily mean anomaly series
(Gunnedah 1969-1995, 31.02$^\circ$ south, 150.27$^\circ$ east). Gray lines
indicate scaling with an exponent $\delta=0.75\pm0.01$, dashed line illustrates
the slope for an uncorrelated process.}
\label{fig1}
\end{figure*}

We consider a fluctuating time series $x_i$, ($i=1,...,N$) sampled
at equidistant times $i\Delta t$. We assume that $x_i$ are
increments of a random walk process around the average
$\langle x\rangle=N^{-1}\sum_{i=1}^N x_i=0$, thus the ``trajectory''
or ``profile'' of the signal is given by integration as
\begin{equation}
y_j=\sum_{i=1}^j x_i \enspace.
\label{Eq:def}
\end{equation}
We divide the profile into nonoverlapping segments of equal length $n$
indexed by $k=1,...,[N/n]$.
In each segment, the local trend is fitted by a polynomial
$f_k^{(p)}(j)$ of order-$p$  [see Fig.~\ref{fig1}(a)],
and the profile is detrended by subtracting this local fit: $z^{(p)}_j=
y_j-f_k^{(p)}(j)$, $j=1,...,N$.
A possible measure of fluctuations can be given by the root mean
square
\begin{equation}
F_p(n) = \sqrt{\frac{1}{n[N/n]}\sum_{j=1}^{n[N/n]} \left(z_j^{(p)}\right)^2}
\label{Eq:rms}
\end{equation}
for a given segment length $n$. A power-law relationship
between $F_p(n)$ and $n$ indicates scaling with an exponent
$\delta$ (DFA$p$ exponent):
\begin{equation}
F_p(n) \sim n^\delta \enspace.
\label{Eq:alph}
\end{equation}
Notice that such a process has a power-law autocorrelation function
\begin{equation}
C(\tau)=\langle x_jx_{j+\tau}\rangle\sim \tau^{-\alpha} \enspace,
\label{Eq:auto}
\end{equation}
where $0<\alpha<1$, and the relationship between the correlation exponents is
(Koscielny-Bunde et al.~1998; Talkner and Weber 2000)
\begin{equation}
\alpha = 2(1-\delta)\enspace .
\label{Eq:rel}
\end{equation}
Consequently, long-memory (persistent) processes are characterized by a DFA
exponent $\delta > 0.5$, uncorrelated time series (e.g., pure random walk) obey
$\delta = 0.5$, antipersistent signals exhibiting negative long range correlations
have $\delta<0.5$. (Here the terms ``persistent'' and ``antipersistent'' are used
in the sense that an increasing trend in the past implies an increasing or decreasing trend in
the future, thus it slightly differs from ``persistence'' in climatology defined as
the continuance of a specific pattern.)

Fig.~\ref{fig1}(b) shows typical results  for a particular station
(Gunnedah). As a first step of the analysis, the annual cycle is removed from the raw
data $T_i$ (daily mean temperatures) by computing the temperature anomaly series
\begin{equation}
x_i=T_i-\langle T_i\rangle _d \enspace,
\label{Eq:anom}
\end{equation}
where $\langle \rangle _d$ denotes the long-time average value for the given calendar
day. The curves in Fig.~\ref{fig1}(b) represent the results for DFA1, DFA2 and DFA3
evaluation, where local trends were removed by linear, second-, and third-order polynomials.
The low noise level is due to the standard ``sliding window'' technique,
where local trend removal and variance computation for a given time window $n$
were performed at each possible starting value $i=1,...,N-n$.
The shapes are practically the same indicating that there is no significant trend in the
temperature anomaly series (DFA$p$ would filter out an overall polynomial trend of
order $p-1$ resulting in changing asymptotic slopes). Note that this observation does
not contradict recent studies on global warming for the same
geographic area, because relatively weak trends are easily masked by the
variability of daily average temperatures, thus  careful statistical
analyses of extremes are proven more convincing (Easterling et al. 1997; Manton et al. 2001;
Frich et al. 2002).
The gradual downward
shift (decreased overall variance for a given segment length $n$) is a simple consequence
of that local trends are better approximated by polynomials of higher order [c.f. Fig.~\ref{fig1}(a)].
For short times ($n<20$ day), the curves exhibit a gradually decreasing slope to a well
defined asymptotic value $\delta=0.75\pm 0.01$ for the particular case, the statistics
breaks down for very large segment lengths ($n > 2000$ day $\approx$ one fifths of the
total length).

Systematic analyses by Hu et al.~(2001), Kantelhardt  et al.~(2001) and
Chen et al.~(2002) with various synthetic time series revealed many useful
details on the efficiency of the method.
One of their main findings  is that DFA results
for signals with different correlation properties and background trends
can be fully explained by the assumption of variance superposition.
This is essential because the correlation exponent is almost never constant
for any real data, crossover(s) can usually arise from a change in the
correlation properties at different time scales or as a consequence of trends.
Other types of  nonstationarities are missing segments in records (very common),
contamination with random spikes, or signals with different local behavior (different
variance or local correlations). From our point of view, the most important cases
are the following (Hu et al.~2001; Kantelhardt  et al.~2001; Chen et al.~2002):
\begin{enumerate}
\item {\it   Long range correlated process with additive random noise.} The typical DFA curve
has a crossover from a slope  $\delta= 0.5$ (noise dominated part) to a different asymptotic
value. The crossover time depends primarily on the variance of the random noise (large amplitude
- later crossover) and on the correlation exponent of the underlying process (large exponent - earlier
crossover).
\item {\it   Noise with sinusoidal trend.} A pure sinusoidal signal has a DFA$p$ curve of initial slope $\delta =p+1$
with a crossover to $\delta =0$. (Obviously, the variance is limited by the total amplitude in time windows  larger
than the half  period of the harmonic signal.) Together with noise, the resulting DFA$p$ curve is a superposition
of both ingredients.
\item{\it   Signals with segments removed.} Surprisingly, scaling of correlated data series ($\delta>0.5$)
is not affected by randomly cutting out segments and stitching together the remaining parts, even when
50\% of the points are removed (Chen et al.~2002). Nevertheless the applicability of this result
for temperature records with long missing segments is limited, because the removal of the annual cycle
can be problematic.
\item {\it   Signals with different local correlations.} In general, when random parts of a correlated
data series is replaced by segments from another series of different correlation exponent, the behavior
is dominated by the segments exhibiting higher positive correlations (Chen et al.~2002).
 However, there is a wide
transition regime with a nontrivial exponent, thus a ``true'' asymptotic behavior could be observed
at very long time series only, especially when the difference between the exponents is small.
We will return to the relevance of this point in Section 4 (Discussion).
\end{enumerate}
Finally, we emphasize that we observed weak overall trends in the longest temperature anomaly
series (Sydney, Melbourne, Adelaide) which might be attributed to urbanization or global warming,
but this question is beyond the scope of the present work. Even for these cases, DFA2
or higher order exponents
could not be distinguished from DFA1 values within the fitting error.

\section{Results}

We analyzed 61 daily maximum, minimum and mean temperature series for Australian stations,
and 18 daily mean temperature records for Hungary. We found
asymptotic long range correlations in the range 30-1800 days for each case,
 the fitting regime of power-law behavior extends up to 10 years for the longest
records. In strong contrast to earlier claims on a universal exponent value
(Koscielny-Bunde et al.~1998; Govindan et al.~2002),
we found pronounced station dependence.

\subsection{Correlation exponent}

\begin{figure*}
\begin{center}
\rotatebox{90}{\resizebox{0.6\textwidth}{!}{\includegraphics{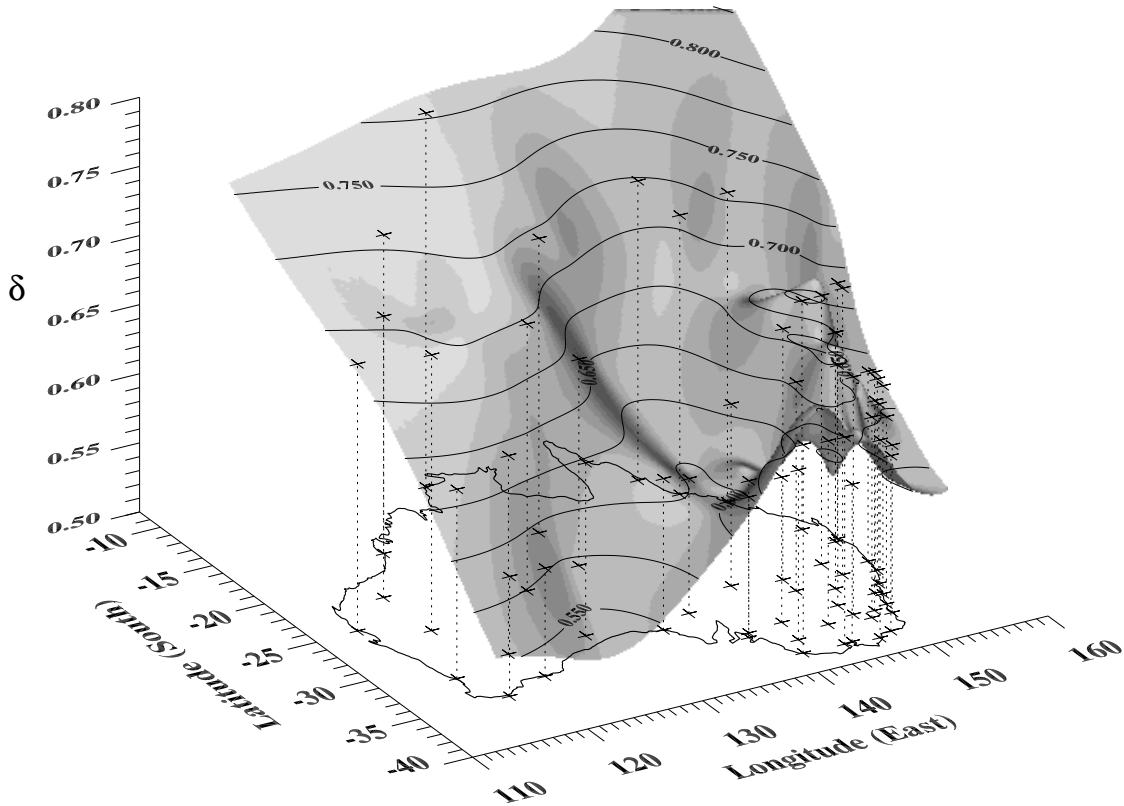}}}
\end{center}
\caption{Correlation exponent $\delta$ for daily mean temperature anomalies
at the continental stations (48 altogether) as a function of
geographic location. The contour of Australia is also indicated. (Error bars are shown
in Fig.~\ref{figsum}.)}
\label{figsurf}
\end{figure*}

Figure \ref{figsurf} illustrates how the value of correlation exponent $\delta$ is connected
with the geographic location of the station over Australia. Here DFA analysis was performed for 48 daily
mean temperature anomaly series, the behavior of daily minima and maxima is practically
the same. Two tendencies can be resolved in Fig.~\ref{figsurf}. Firstly, the general trend
is a decrease of exponents with decreasing latitude. Indeed, the contours of the fitted
surface are almost parallel with the lines of latitude.
This tendency is in complete agreement found by Tsonis {\it  et. al.}~(1999) for
500-hPa height anomalies in the northern hemisphere, which indicates that the
key factor is the distance from the equator.
Secondly, there is a  hump
over the south-eastern part of the continent which correlates well with the location of the
highest mountain range (Australian Alps). We do not want to overemphasize this
observation, nevertheless Weber and Talkner (2001) found also higher exponent values for
mountain weather stations.
%Note that a similar evaluation for Hungary is not possible, simply because its south-north
%extension is 2$^\circ$47' only (see Fig.~\ref{figsum}).

\begin{figure*}
\begin{center}
\resizebox{0.99\textwidth}{!}{\includegraphics{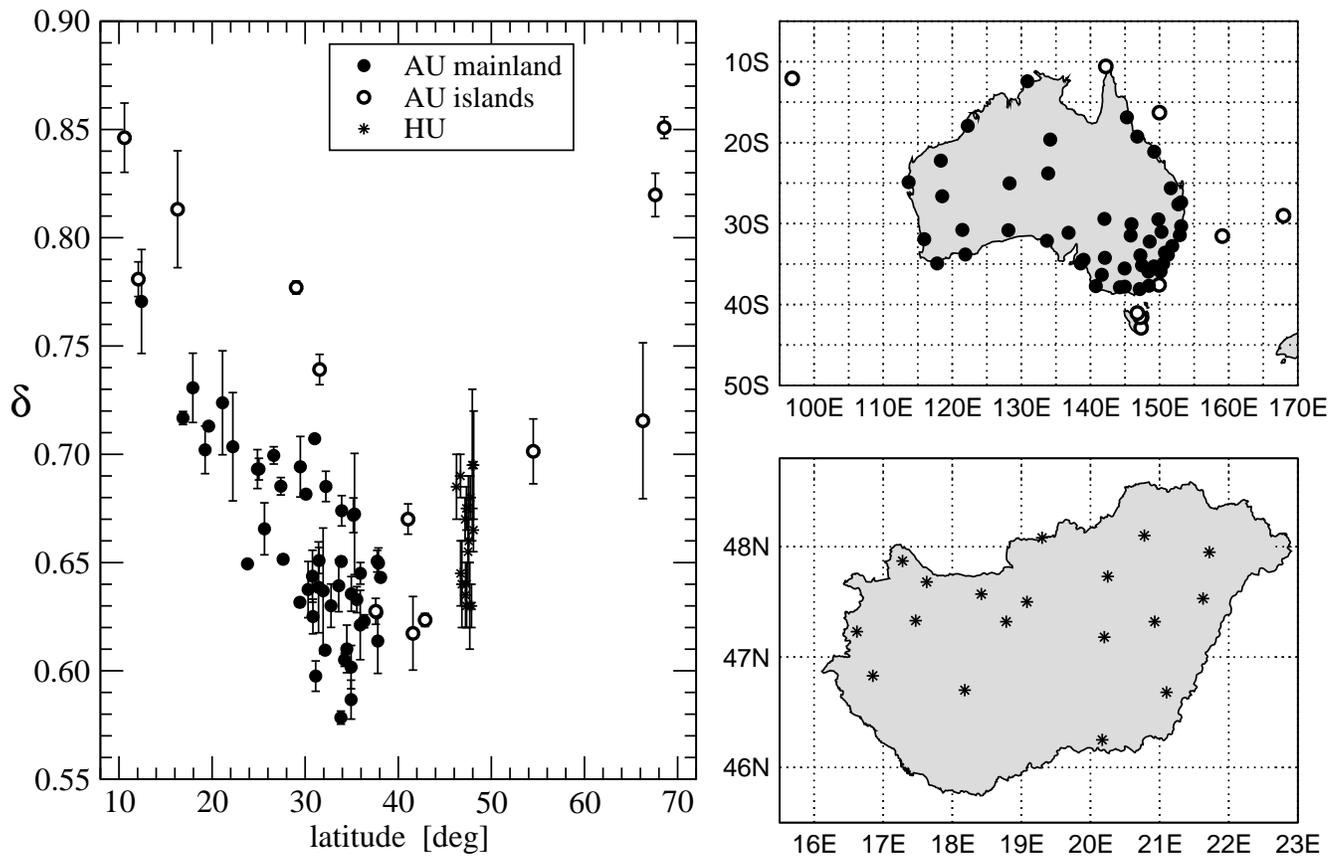}}
\end{center}
\caption{Correlation exponent $\delta$ for daily mean temperature anomalies as a function
of geographic latitude. Heavy dots: Australian continent; empty circles: islands (note that
locations south from the line 50$^\circ$S are not indicated in the map);
stars: Hungarian stations. The legend box does not hide data.}
\label{figsum}
\end{figure*}

The  decreasing tendency over the Australian continent is clear also in Fig.~\ref{figsum},
where the same exponent $\delta$
is plotted as a function of latitude for every analyzed stations. Error bars are obtained from
fits to different segments for the scaling regime of DFA curves.
A comparison with Hungarian
data (stars in Fig.~\ref{figsum}) indicate another aspect of the
 the lack of universality:  geographic latitude is not a dominant factor over
 the Carpathian Basin. Note that a full symmetry of the northern and
southern hemispheres is not expected, also the differences of climates should be reflected
in differing correlation properties. As for the islands, Weber and Talkner (2001), and
Monetti {\it  et. al.}~(2003) already reported on systematically higher level of correlations
for maritime stations and sea surface temperatures, which is in agreement with our
results (see Fig.~\ref{figsum}). It is interesting that the trend for islands is also decreasing
upto 50$^\circ$S. The number of cases  below this latitude is not enough to
conclude whether the trend is reversed there, or what we see is just
localized anomaly. These four stations
(in latitudinal order: Macquarie Island, Casey, Mawson, and Davis) belong essentially to Antarctic
climate.

It is remarkable that our main result (decreasing exponents with increasing distance from the equator)
is not visible on the map given by Fraedrich and Blender (2003), which might be a consequence
of their much lower spatial resolution over Australia.

\begin{figure*}
\begin{center}
\resizebox{0.9\textwidth}{!}{\includegraphics{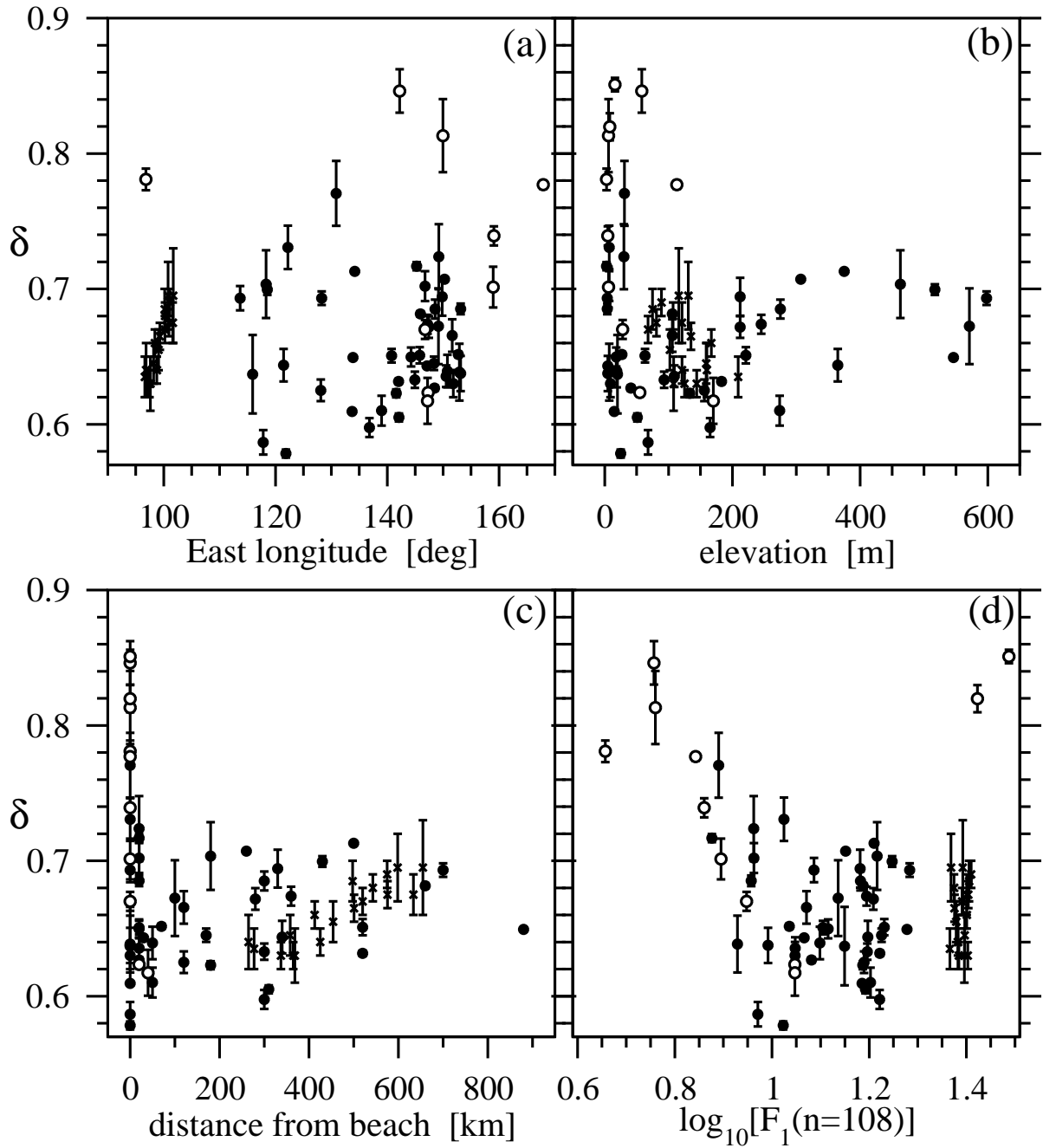}}
\end{center}
\caption{Scatter plot for the correlation exponent $\delta$, and {\bf (a)} the longitude
(values for Hungary are shifted by 80$^\circ$), {\bf (b)} the elevation, {\bf (c)} the distance
from the closest seashore (Trieste for Hungarian stations), and {\bf (d)}
the logarithm of average variance of fluctuations
for  time window $n=108$ days. Heavy dots: Australian mainland; empty circles: islands;
stars: Hungarian stations.}
\label{figstat}
\end{figure*}

In order to reveal another possible determining factors, we show in Fig.~\ref{figstat}
scatter plots for the exponent value $\delta$ and different geographic parameters.
Hungarian data in Fig.~\ref{figstat}a (stars) indicate that longitude is a dominating factor over this
region, unfortunately the narrow spatial coverage restricts an unambiguous conclusion.
This behavior is reflected also in Fig.~\ref{figstat}c, but this is an obvious consequence
of geographic locations: whatever is the reference point (Mediterranean Sea or Atlantic Ocean),
the distance from the shore is increasing as the longitude increases.

As for Australia, Fig.~\ref{figstat}a can be explained by accepting that latitude is the
determining factor, and there is no correlation with longitude. In this case we expect a
distribution reflecting somehow the shape of the continent. Indeed, the higher density
of stations at the south-west corner is seemed to be projected  in Fig.~\ref{figstat}a.
Cumulated exponent values show weak correlation with the elevation, see Fig.~\ref{figstat}b.
Note that this does not contradict to the interpretation we gave for the ridge
in Fig.~\ref{figsurf}, because this scatter plot smears differences in longitude and latitude
of stations. As for the distance from the oceans (Fig.~\ref{figstat}c), the dependence
is week again. High exponent values can be associated with islands, but there is no
significant drop as we move away from the shore. This does not mean that the distance
from large water masses can not contribute to the degree of correlation, but the apparent
effect is weak over Australia.

Fig.~\ref{figstat}d shows the scatter plot for $\delta$ and the DFA variance at a fixed
temporal window size, but let us return to this point after describing the general
properties of variance itself, in the next Subsection.

\subsection{Variance}

\begin{figure*}
\begin{center}
\rotatebox{90}{\resizebox{0.3\textwidth}{!}{\includegraphics{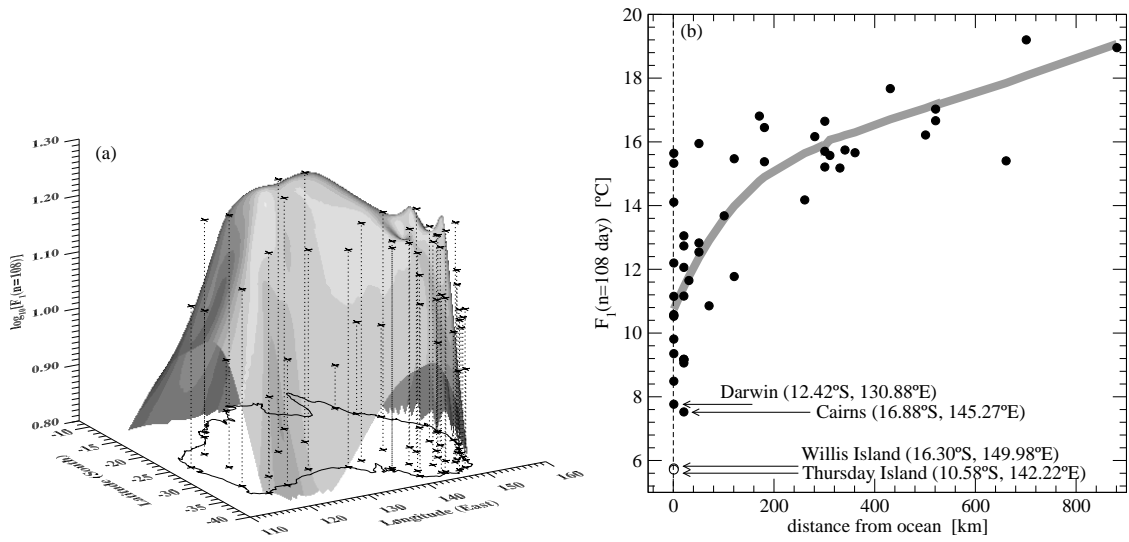}}}
\resizebox{0.4\textwidth}{!}{\includegraphics{KJ2fig5b.eps}}
\end{center}
\caption{Average variance of fluctuations for daily mean temperature anomalies
for  time window $n=108$ day. (a): $Log_{10}[F_1(n=108)]$ (see Eq.~\ref{Eq:rms})
at the location of stations on the Australian mainland.
(b): $F_1(n=108)$ as a function of distance from the ocean.
(The gray line only guides the eye.) Stations of smallest variance are identified [see
Section 3(c)].}
\label{figvari}
\end{figure*}

In order to characterize the average variance of fluctuations, we plotted
$F_1(n=108)$ (see Eq.~\ref{Eq:rms}) for daily mean temperature anomalies
in Fig.~\ref{figvari} as a function of geographic location. Daily minima and maxima
show the same behavior (but not necessarily the same numerical value), again. The
particular time lag might seem to be somewhat arbitrary, however any other choice
in the range $100<n<500$ resulted in  the same statistics, apart from numerical
values. Too large time lags are not suitable, because the DFA curves for different
stations have different slopes,
and crossing confounds the relationship between stations.
The general trend for variance over the continent is not surprising: the proximity of ocean damps temperature
fluctuations, which is a well known and well documented fact.

\begin{figure*}
\begin{center}
\resizebox{0.45\textwidth}{!}{\includegraphics{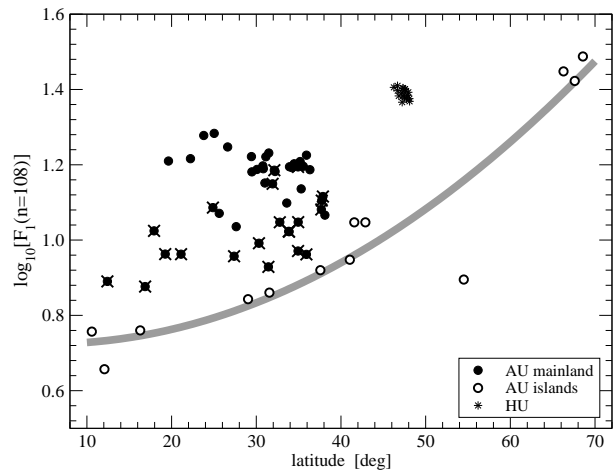}}
\end{center}
\caption{
Logarithm of average variance of fluctuations for daily mean temperature anomalies $F_1$
for  time window $n=108$. Solid circles: Australian mainland; solid circles with crosses:
Australian shore (distance from the ocean is less than 20 km); empty circles: islands; stars:
Hungarian stations. The gray line only guides the eye.}
\label{figvarisum}
\end{figure*}

Figure \ref{figvarisum} shows the variance statistics for every stations as a function of geographic
latitude. There is a trend of increasing variance with increasing distance from
the equator, especially for the islands. This tendency can be revealed also by marking continental
stations on the shore (solid circles with crosses in Fig.~\ref{figvarisum}), however the dispersion
is rather large. Variances far from the oceans are systematically increased both for Australia
and Hungary (Fig.~\ref{figvarisum}).

The correlation plot between average variance and DFA exponents is shown in Fig.~\ref{figstat}d.
Based on Fig.~\ref{figsum}a and Fig.~\ref{figvarisum}, we expect that stations
far away from the equator obey a relatively low exponent value  but large variance,
this negative correlation is clearly visible in Fig.~\ref{figstat}d.
The scatter for continental stations is pretty large, thus one-to-one coupling remains hidden
(if it exists at all) by
other modifying factors, such as the distance from ocean or elevation.

\subsection{Detection of Tropospheric Biennial Oscillation}

\begin{figure*}
\begin{center}
\resizebox{0.95\textwidth}{!}{\includegraphics{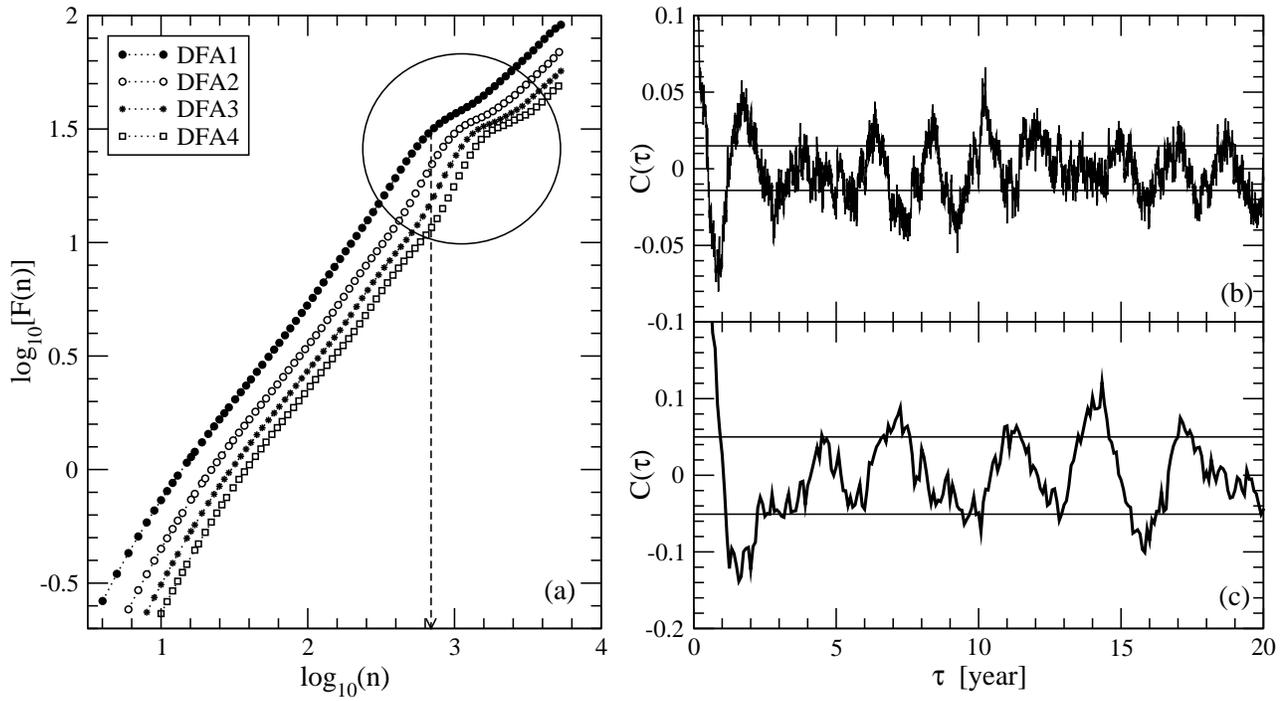}}
\end{center}
\caption{(a) DFA analysis for Willis Island (1939-1999, 16.30$^\circ$S, 149.98$^\circ$E).
Circle indicates the kink in the curves characteristic for a quasi-periodic background
cycle of $\sim 1.9$ years (vertical dashed line).
(b) Autocorrelation function for the same series (solid line), the lag $\tau$ in units of years.
(c) Autocorrelation for the Southern Oscillation Index (1866-1995)
from the University of East Anglia
(ftp://daac.gsfc.nasa.gov/data/inter\_disc/surf\_temp\_press/soi),
$\tau$ in units of years.
Thin lines indicate  95\% confidence level for both autocorrelation functions.}
\label{figsoi}
\end{figure*}

In a few cases, DFA curves exhibited characteristic kinks shown in Fig.~\ref{figsoi}(a),
which became more pronounced at higher order local detrending. This behavior is
consistent with the presence of a periodic sinusoidal background trend (Hu et al.~2001).
Unfortunately, there is no way to decompose such an underlying component, especially
because it is far from being pure harmonic indicated by the shift of breakpoints [see
Fig.~\ref{figsoi}(a)].

The existence of a slow, smeared oscillation was actually clear for three stations, namely
Cairns (16.88$^\circ$S, 145.27$^\circ$E),
Thursday Island (10.58$^\circ$S, 142.21$^\circ$E), and
Wil\-lis Island (16.30$^\circ$S, 149.98$^\circ$E); a further time series
(Darwin, 12.42$^\circ$S, 130.88$^\circ$E) indicated a weak signal of similar behavior.
For these cases, even the standard autocorrelation analysis [Fig.~\ref{figsoi}(b)]
revealed statistically significant periodicity of $1.9\pm 0.2$ year
(note that this is an average considering also higher harmonics).
The very stations are located at the north-eastern
edge of the continent (Fig.~\ref{figsum}), and they exhibit the lowest intrinsic variance,
see Fig.~\ref{figvari}. [Actually, the lowermost variance is found at Cocos (Keeling) Island
(12.07$^\circ$S,  96.83$^\circ$E), c.f.~Fig.~\ref{figvarisum}, but it is situated
far to the west in the Indian Ocean.)
No other stations have shown statistically significant periodic components.

The location of these places suggests that the observed  periodic trend might
be connected with the well-known El-Ni\~no Southern Oscillation (ENSO), that is
the quasi-periodic disruption of the Walker circulation forced by a  pressure
gradient over the Pacific. The changes in pressure distribution are measured by
the Southern Oscillation Index (SOI), which is far from being regular
with a period of 2-7 years
(Ropelewski and Jones 1987; Allan {\it  et.~al.}~1991).
A direct comparison of the autocorrelation functions for the daily mean temperature
anomalies at Willis Island [Fig.~\ref{figsoi}(b)] and for the SOI
signal [Fig.~\ref{figsoi}(c)] shows immediately that the two oscillations
are not strongly coupled.

Biennial signals in rainfall, sea-level pressure, sea-surface temperature and other
climate elements in the tropical Indian and western Pacific regions have been known
and studied for years (e.g.,Trenberth 1975; Meehl 1987; 1997).
The phenomenon is termed as Tropospheric Biennial Oscillation (TBO).
This TBO has an irregular period of 2-3 years and shows eastward movement.
Several theories and hypotheses
have been proposed, however the exact mechanism  is not clearly known.
A strong biennial component is seen in the ENSO phenomena (Ropelewski {\it  et.~al.}~1992)
and it is accepted  to be an integral part of the Asia-Pacific climate.
Since the slow oscillation observed for the stations above has a characteristic
period of two years [see Fig.~\ref{figsoi}(b)], we can consider it as a fingerprint
of TBO. As far as we know, no other daily surface data series of similar length
have shown such markedly the presence of two-year oscillation.

\section{Discussion}
\renewcommand{\labelenumi}{(\roman{enumi})}

Our main findings can be summarized as follows.
\begin{enumerate}
\item Daily temperature values have  significant positive long range correlations extending for
several years. There is no sign of breakdown of scaling behavior even for the longest time series.
\item The ``degree'' of correlations, i.e., the value of correlation exponent is not universal, it depends
on the geographic location.
\item There is a general tendency of decreasing exponent value with increasing distance from the equator
 over Australia.
\item The variance of fluctuations has an opposite trend: the larger the distance from the equator,
the larger the intrinsic variance.
\end{enumerate}

Let us comment shortly points (ii) and (iii). As for the lack of universality, (ii),  an opposite conclusion
could be drawn from DFA analysis of a few time series only
(Koscielny-Bunde et al.~1998; Govindan et al.~2002).
Similar tests by Weber and Talkner (2001),
Monetti {\it  et. al.}~(2002),  Fraedrich and Blender (2003),
and especially Eichner {\it et al.}~(2003)
do indicate significant differences as a function of
climatic parameters of the stations. Different exponent values for different places
do not exclude the existence of a universal mechanism behind the long memory effects,
such as coupling with the oceans.
Nevertheless any speculation on such a mechanism should be preceded
by further analyses of asymptotic correlations with the highest possible spatial
resolution.

The decreasing tendency of correlation exponents over Australia
is in complete agreement with the results  by Tsonis {\it  et. al.}~(1999) for
500-hPa height anomalies in the northern hemisphere. Their fluctuation analysis
is essentially identical with a DFA1 method, even  exponent values can
 be compared. They attribute a decreasing degree of correlations to
an increasingly baroclinic nature of the dynamics as one progresses from
the subtropics through the midlatitudes. More baroclinicity results
in more power to processes of small spatial and temporal scales.
Our results support this picture in the following way. Whatever is the source
of a decreased correlation exponent at higher latitudes, it can not be decomposed
to a core dynamics and an additive random noise. This is because the DFA method
very effectively identifies such a process. Tuning of the exponent value is possible, e.g.,
by replacing random parts of a correlated signal with segments from another series of different
correlation properties (see Section 1).
In this respect, a dynamics of  increased baroclinicity means that an underlying correlated
process can be more often interrupted by short-memory, small scale processes
yielding  a decreased effective correlation exponent.

\section*{Acknowledgements}

We would like to thank Tam\'as T\'el  for useful
recommendations  and for his
permanent encouragement.
This work was supported by the Hungarian Science Foundation (OTKA) under
Grant Nos. T032423 and T032437.

\newpage

\end{document}